\documentclass[fleqn,usenatbib]{mnras}
\usepackage{float}
\usepackage{newtxtext,newtxmath}
\usepackage{arydshln}
\usepackage[T1]{fontenc}
\DeclareRobustCommand{\VAN}[3]{#2}
\let\VANthebibliography\thebibliography
\def\thebibliography{\DeclareRobustCommand{\VAN}[3]{##3}\VANthebibliography}
\usepackage{graphicx}	
\usepackage{amsmath}	

\title[Is EZ CMa a fast-precessing binary?]{Using CHIRON Spectroscopy to Test the Hypothesis of a Precessing Orbit for the WN4 star EZ CMa}
\author[K. DG. Barclay et al.]{Krister DG. Barclay$^{1,2}$\thanks{E-mail: krister.barclay@stonybrook.edu},
Sophie Rosu$^3$\thanks{E-mail: srosu@kth.se},
Noel D. Richardson$^{1}$\thanks{E-mail: noel.richardson@erau.edu},
Andr\'e-Nicolas Chen\'e$^4$,
\newauthor 
Nicole St-Louis$^5$, 
Richard Ignace$^6$,
and Anthony F. J. Moffat$^5$
\\
$^{1}$Physics and Astronomy, Embry-Riddle Aeronautical University, 3700 Willow Creek Road, Prescott, AZ 86301, USA\\
$^2$Department of Physics \& Astronomy, Stony Brook University, Stony Brook, NY 11794-3800, USA \\
$^3$Department of Physics, KTH Royal Institute of Technology, The Oskar Klein Centre, AlbaNova, SE-106 91 Stockholm, Sweden\\
$^4$ US-ELTP/NSF's NOIRLab, 670 N. A'ohoku Place, Hilo, HI, 96720, USA\\
$^5$D\'epartement de physique, Universit\'e de Montr\'eal, Complexe des Sciences, 1375 Avenue Th\'er\`ese-Lavoie-Roux, Montr\'eal (Qc), H2V 0B3, Canada \\
$^6$Department of Physics \& Astronomy, East Tennessee State University, Johnson City, TN 37614, USA \\
}

\begin{document}
\label{firstpage}
\pagerange{\pageref{firstpage}--\pageref{lastpage}}
\maketitle

\begin{abstract}
The bright WN4 star EZ CMa exhibits a 3.77 day periodicity in photometry, spectroscopy, and polarimetry but the variations in the measurements are not strictly phase-locked, exhibiting changes in reference times, amplitudes, and the shape of the variability happening over times as short as a few weeks. Recently, 137 days of contiguous, variable photometry from BRITE-Constellation was interpreted as caused either by large-scale dense wind structures modulated by rotation, or by a fast-precessing binary having a slightly shorter 3.626 day orbital period and a fast apsidal motion rate of $1315^\circ\,\text{yr}^{-1}$. We aim at testing the latter hypothesis through analysis of spectroscopy and focus on the N\,{\sc v} $\lambda\,4945$ line. We derive an orbital solution for the system and reject the 3.626 day period to represent the variations in the radial velocities of EZ CMa. An orbital solution with an orbital period of 3.77 days was obtained but at the cost of an extremely high and thus improbable apsidal motion rate. Our best orbital solution yields a period of $3.751\pm0.001$\,days with no apsidal motion. We place our results in the context of other variability studies and system properties.
While we cannot fully reject the precessing binary model, we find that the corotating interaction region (CIR) hypothesis is better supported by these and other data through qualitative models of CIRs.

\end{abstract}


\section{Introduction}
Classical Wolf-Rayet (WR) stars represent an evolved stage of massive-star evolution, where the progenitor O stars must have a mass greater than $\sim 25 M_\odot$. They have dense winds (with a mass-loss rate $\dot{M}\,\sim\,10^{-5}\,M_\odot\,{\rm yr}^{-1})$ that have high terminal velocities ($v_\infty\,\sim\,2000\,{\rm km\,s}^{-1}$) and are observed to lack hydrogen in their spectra \citep[see a review in ][]{2007ARA&A..45..177C}. These stars are classified based on their abundances as nitrogen-rich (WN), carbon-rich (WC), or oxygen-rich (WO). Recent analyses of Galactic WN and WC stars showed that these stars are often found in binaries: about 56\% of WN stars \citep{2022arXiv220412518D} and at least 72\% of WC stars \citep{2020A&A...641A..26D}. Amongst the WN binaries -- or candidates --, the bright EZ CMa (EZ Canis Majoris = WR6, HD\,50896) Wolf-Rayet star is an unusual test case: depending on the interpretation of its variability, it could represent a rotating single or binary star. 

With $V=7.0$ mag, WR6 is the 6th brightest WR star in the sky. It is a hot luminous WR star of the nitrogen sequence with sub-type WN4b, where the `b' indicates `broadline' \citep{1996MNRAS.281..163S}. WR6 is surrounded by a prominent wind-swept ring nebula S308, whose radial velocity (RV) along with its relatively large separation from the Galactic plane and stellar proper motion, suggest that WR6 is a runaway star, an interpretation somewhat tempered by the enhanced warp and thickness of the Galactic disk in the outer regions where EZ CMa is located \citep[e.g.,][]{2020RNAAS...4..213G}. Not being an obvious WR+O binary, WR6 was often considered a prototype single star. However, its strong variability soon became apparent, making it less appropriate as such an object. \citet{1980ApJ...239..607F} first discovered a periodicity in the variations, with $P=3.763\pm0.002$\,days, though with clear and strong epoch dependency, especially for the amplitude and zero-point of its phased (mainly) spectral variability. These authors claimed that the best explanation was in terms of a runaway binary with a neutron star companion to the bright WR component which formed during the slingshot runaway process of a supernova leading to the neutron star. However, X-ray data do not support the presence of such a compact companion \citep{1988MNRAS.234..783S,1989ApJ...347..409P, 2012ApJ...747L..25O, 2015ApJ...815...29H}. Therefore, another process must have led to the runaway status, for instance the star is located higher than usual in the external Galactic- plane warp and larger width, the neutron star went off flying in the opposite direction, or the star was catapulted out of a nearby forming star cluster/association such as Cr\,121.

Later, \citet{1992ApJ...397..277R} summarized the optical variability based on repeated precision photometry, spectroscopy, and polarimetry. Of particular interest was the photometry, mainly because of the large number of observations on different timescales, covering a decade at that time. These photometric observations reported by \citet{1992ApJ...397..277R} confirmed the strong epoch dependency from essentially no variability at some epochs to a coherent $0.1$ mag amplitude at others. A Fourier analysis of the photometric amplitude led to two marginally possible coherencies for $P=101$\,days and $651$\,days, with the latter somewhat more significant. The linear polarimetry revealed marginally possible binary-orbit fits with epoch-dependent varying eccentricity but little variation in $\omega$, the longitude of periastron. The circular-polarisation has not shown any clear variation, eliminating the possibility of a rotating global strong magnetic field with cyclotron emission from non-relativistic electrons, as seen in magnetic cataclysmic variables (polars) with magnetic fields of megagauss strength \citep[e.g.,][]{2012ASPC..465...92D, 2016MNRAS.458.3381H}. This result was later confirmed by \citet{2013ApJ...764..171D}. The spectroscopy showed little coherency among different lines and between different epochs.
The 3.77 day period examined by \citet{1997ApJ...482..470M} showed it to be consistent over large data sets, even if the variability is epoch dependent.
Overall, the only viable explanations for the 3.77 day period are a binary with a low-mass companion (favoring a neutron star, although not supported in some X-ray analyses) or a single rotating star with large-scale epoch-dependent inhomogeneities. Also, pulsations are highly unlikely as an explanation for the periodic behavior of EZ CMa given the coherence in the variability seen in other pulsating WR stars such as WR 123 \citep{2005ApJ...634L.109L}. 

Another important optical spectral study was later carried out by \citet{1997ApJ...482..470M,1998ApJ...498..413M}, showing a clear preference for rotating large density structures in the wind of the single star, likely stemming from hot and/or magnetic spots on the stellar hydrostatic surface. These would then be identified with corotating interaction regions (CIRs) arising from bright spots on the stellar surface, as seen in essentially all O stars and modelled by \citet{1996ApJ...462..469C}. There is no obvious reason why a similar phenomenon should not be seen among WR stars, despite their hidden inner dense winds, thus reducing the frequency of detection, but EZ CMa may be the clearest case. CIR modelling by \citet{2002A&A...395..209D} showed a good qualitative match with the dynamic spectra observed by \citet{1997ApJ...482..470M,1998ApJ...498..413M} for EZ CMa, assuming there were several CIRs per rotation, that were spaced out randomly in azimuth \citep{2011ApJ...735...34C} such as seen with two other nitrogen-rich WR stars, namely WR 1 \citep{2010ApJ...716..929C} and WR 134 \citep{1994Ap&SS.221..155M, 2016MNRAS.460.3407A}. 

More recently, \citet{2019A&A...624L...3S} and \citet{2020A&A...639A..18K} revived the binary option in a new proposed complex scenario of a massive third star causing an inner binary to precess. The authors derived a synodic period of  3.626\,days, an eccentricity of $0.102\pm0.010$, and an apsidal motion rate of $-15.6^{\circ} {\rm orbit}^{-1}$ for the inner binary. This idea stemmed from the best-ever optical light-curve (LC) obtained so far for WR6, by one of the BRITE nanosatellites over a non-stop 4.5 month-long period. Although a more detailed analysis of this BRITE LC along with simultaneous spectroscopy is pending (St-Louis et al., in prep.), a preliminary report of these data was given in \citet{2018pas8.conf...37M} and \citet{2020svos.conf..423S} in terms of CIRs. One can note that TESS data are now also available, though they only cover a month at a time, which is not as complete as the BRITE light curves, despite the higher precision and cadence of the former. Such longer data strings are essential to deal with the epoch dependency. 

In this paper, we analyse regular, repeated precision newly acquired optical spectroscopy with the CHIRON spectrograph in order to narrow down the possible scenario to explain EZ CMa’s periodic variability. Section\,\ref{sect:observations} presents our spectroscopic data set and reduction process. The measurements of the line shapes and variability are presented in Sect.\,\ref{sect:variability}. In Sect.\,\ref{sect:apsidal_motion} we derive an orbital solution for the system accounting for apsidal motion through the analysis of the RVs. We discuss our results in the context of other observations of the star in Sect.\,\ref{sect:discussion}, and conclude in Sect.\,\ref{sect:conclusion}.

\section{Observations and Data Reduction}
\label{sect:observations}
In order to test the binary hypothesis for the nature of EZ CMa, an observational campaign was instigated with the CTIO 1.5m telescope and the CHIRON spectrograph \citep{2013PASP..125.1336T} during the time period spanning from November 2020 to March 2021. Each observing run was designed to give a two-week long observation set with one spectrum per night so that radial velocities could be measured and such that the 3.77 day period had reasonable phase coverage.  As a result of these constraints and available telescope time, a total of 54 spectra were collected over four epochs. We define each section of data to represent one of these observing runs, where section 1 was between 2020 November 17--30, section 2 was between 2020 December 07--19, section 3 was between 2021 January 04--17, and section 4 was between 2021 March 04--18.

Each observation consisted of a 400 second exposure that provided a spectrum with a signal-to-noise ratio of 150--200 in the continuum. The resolving power is $R\sim 28\,000$ across the spectrum. Beyond the region of $\sim$\,6500\,\AA, the \'echelle orders no longer overlap resulting in gaps in the wavelength coverage. Wavelength calibration was done for each night through the pipeline process which has been shown to be stable across an entire observational night \citep{2013PASP..125.1336T}. The signal-to-noise peaks around 5500\,\AA, and could vary by a factor of a few across the chip. At the longest wavelengths near 8000\,\AA, the CHIRON throughput is only $\sim10\%$ of that at the peak.

\begin{figure}
\includegraphics[clip=true, trim = 20 0 40 30,width=\linewidth]{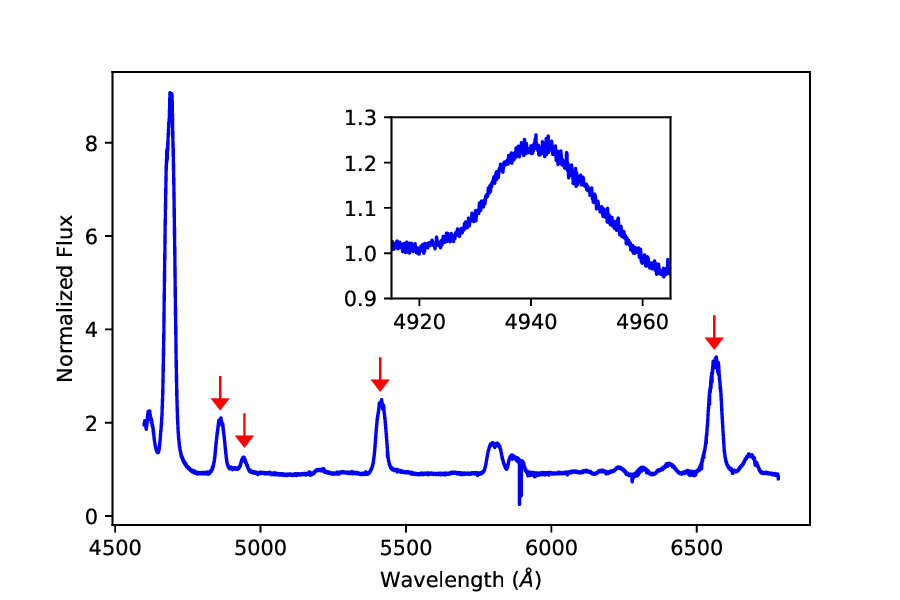}
\caption{An example spectrum from the first night of observation on 2020 November 17. The lines of interest (indicated by the red arrows from left to right) are He\,{\sc ii} $\lambda\,4861$, N\,{\sc v} $\lambda\,4945$, He\,{\sc ii} $\lambda\,5412$, and He\,{\sc ii} $\lambda\,6560$. The inset shows a zoom on the N\,{\sc v} $\lambda\,4945$ line.}
\label{fig:sample spec}
\end{figure}

As described in \citet{2013PASP..125.1336T}, the CHIRON spectrograph is an optical high-resolution spectrometer designed for collecting large amounts of data with a high stability and goal of measuring radial velocities of bright stars. The spectral range for CHIRON is fixed and covers 4550 to 8800\,\AA. The data are run through a pipeline reduction, recently described by \citet{2021AJ....162..176P}, that performs the bias and flat corrections, while providing a wavelength solution with a precision below 1\,km\,s$^{-1}$. We also used the “normal” star HR\,4468 (B9.5V) to fit the blaze function and remove it from our observations, allowing for the data to be compared over multiple \'echelle orders when needed, which was previously done for CHIRON observations of hot stars with broad emission lines \citep{2016MNRAS.461.2540R,2017MNRAS.471.2715R}. An example spectrum of EZ CMa is shown in Fig.\,\ref{fig:sample spec}. 

\section{Measurements of the Line Shapes and Variability} 
\label{sect:variability}
The following regions of interest were investigated closely: for He\,{\sc ii} $\lambda\,4861$, N\,{\sc v} $\lambda4945$, He\,{\sc ii} $\lambda\,5412$, and He\,{\sc ii} $\lambda\,6560$, we constrained our measurements to the regions $4820 - 4900$\,\AA, $4915 - 4965$\,\AA, $5350 - 5550$\,\AA, and $6500 - 6620$\,\AA, respectively. The analysis involved the evaluations of the standardized moments where the equivalent width (EW) is the zeroth moment, the radial velocity (RV) the first, the square of line width the second, skewness (S) the third, and kurtosis (K) the fourth. The EWs were measured from 
\begin{equation}
W_{\lambda} = \int \frac{F_{c} - F_{\lambda}}{F_{c}}d\lambda,
\end{equation}
\noindent where $F_{\lambda}$ is the normalized flux across the wavelength region of interest and $F_{c}$ is the flux of the continuum spectrum across the line, assumed to be equal to unity after normalization.

The $S$ and $K$ measurements were computed using the following equations:
\begin{equation}
S = \frac{\mu_{3}}{\mu_{2}^{\frac{3}{2}}},
\end{equation}
and 
\begin{equation}
K = \frac{\mu_{4}}{\mu_{2}^{2}},
\end{equation}
with 
\begin{equation}
\mu_n = \frac{\sum_{i}(\lambda_{i} - \bar\lambda)^{n}I_{i}}{\sum_{i}I_{i}}
\end{equation}
and 
\begin{equation}
\bar\lambda = \frac{\sum_{i}\lambda_{i}I_{i}}{\sum_{i}I_{i}},
\end{equation}

\noindent where $I_{i}$ is the intensity of the line at the wavelength $\lambda_{i}$. The skewness and kurtosis measurements indicate changes in the line profiles. Skewness measures the symmetry of a line, where a skewness of $0$ indicates a symmetric line, and a positive or negative value indicates that the line is leaning more towards the blue or red of the symmetric normal distribution, although line wings on the red side of WR wind-line profiles will often shift the zero-point to a positive value. Kurtosis gives a measure of whether the normal distribution line is heavy-tailed ($K > 0$) or light-tailed ($K < 0$). The resulting values for the N\,{\sc v} $\lambda\,4945$ line are provided in Table \ref{Table: N V moments}. We adopt $\bar\lambda$ to calculate the radial velocity. The statistical error of this measurement is our reported error bar in Table \ref{Table: N V moments}, but we think this is a lower limit to the actual error, which does not include any stochastic components from the wind variability at the time of the observation.

\section{Test of an Orbital solution}\label{sect:apsidal_motion}
An issue encountered when examining potential orbital variations for EZ CMa is the extremely variable emission-line profiles of WR 6. We here decided to restrict our radial velocity analysis to the sole N\,{\sc v} $\lambda$\,4945 line as the  He\,{\sc ii} lines are formed at much larger radii than the high-ionization N\,{\sc v} line. 
Indeed, in the presence of additional asymmetries in the wind, the lines with smaller formation radii have a much better chance to sample the motion of the star itself rather than of the wind asymmetries. For example, in the case of the bright WR binary $\gamma^2$ Vel, \citet{2017MNRAS.471.2715R} showed that the semi-amplitude of motion of the WR star is related to the upper energy state of the atom/ion used in the measurement.
 
We tested the binary hypothesis through a fit of the radial velocities of EZ CMa. We here explicitly assumed EZ CMa is the primary star of the binary system, and proceeded in the fit of the RVs adopting the formalism already used in previous studies \citep[e.g.,][]{2016A&A...594A..33R, 2020A&A...635A.145R, 2022A&A...664A..98R, 2023MNRAS.521.2988R}.

For each time of observation $t$, we fit the RV data with the following relation 
\begin{equation}
\label{eqn:RVp}
\text{RV}_\text{P}(t) = \gamma_\text{P} + K_\text{P} [\cos(\phi(t)+\omega(t)) +e\cos\omega(t)],
\end{equation}
where $\gamma_\text{P}$, $K_\text{P}$, $e$, and $\omega$ are the primary apparent systemic velocity, the semi-amplitude of the primary RV curve, the eccentricity, and the argument of periastron of the primary orbit, respectively. The true anomaly $\phi$ is inferred from the eccentric anomaly, itself computed through Kepler's equation, which involves both $e$ and the anomalistic orbital period $P_\text{orb}$ of the system. We accounted for the apsidal motion through the variation of $\omega$ with time following the relation 
\begin{equation}
\omega(t) = \omega_0 + \dot\omega(t-T_0),
\end{equation}
where $\dot\omega$ is the apsidal motion rate and $\omega_0$ is the value of $\omega$ at the time of periastron passage $T_0$. The systemic velocity was adjusted so as to minimise the sum of the residuals of the data about the curve given by Eq.\,\eqref{eqn:RVp}.

For the RVs, we adopted error bars four times larger than the error bars quoted in Table\,\ref{Table: N V moments} as we found that our derived error bars were underestimated by a factor of $\sim 4-5$ as judged by the quality of the fit. In this case, the error bars on the RVs amount to $\sim 3-4$\,km\,s$^{-1}$, values which are more coherent with the signal to noise ratio and spectral resolution of the spectroscopic data. 

In a first attempt to derive an orbital solution for the system, we fixed the orbital period to the value of 3.626\,days as suggested by \citet{2019A&A...624L...3S} while leaving all other parameters ($T_0, \omega_0, e, K_\text{P}$, and $\dot\omega$) free. We scanned the 5D parameter space but were unable to find any consistent solution as the RV data were poorly adjusted with a reduced $\chi^2_\nu\sim 100$. This result suggests that an orbital period of 3.626\,days is not appropriate to describe the observed RV variations.

\begin{figure*}
\centering
\includegraphics[clip=true, trim=0 50 20 0,width=1\linewidth]{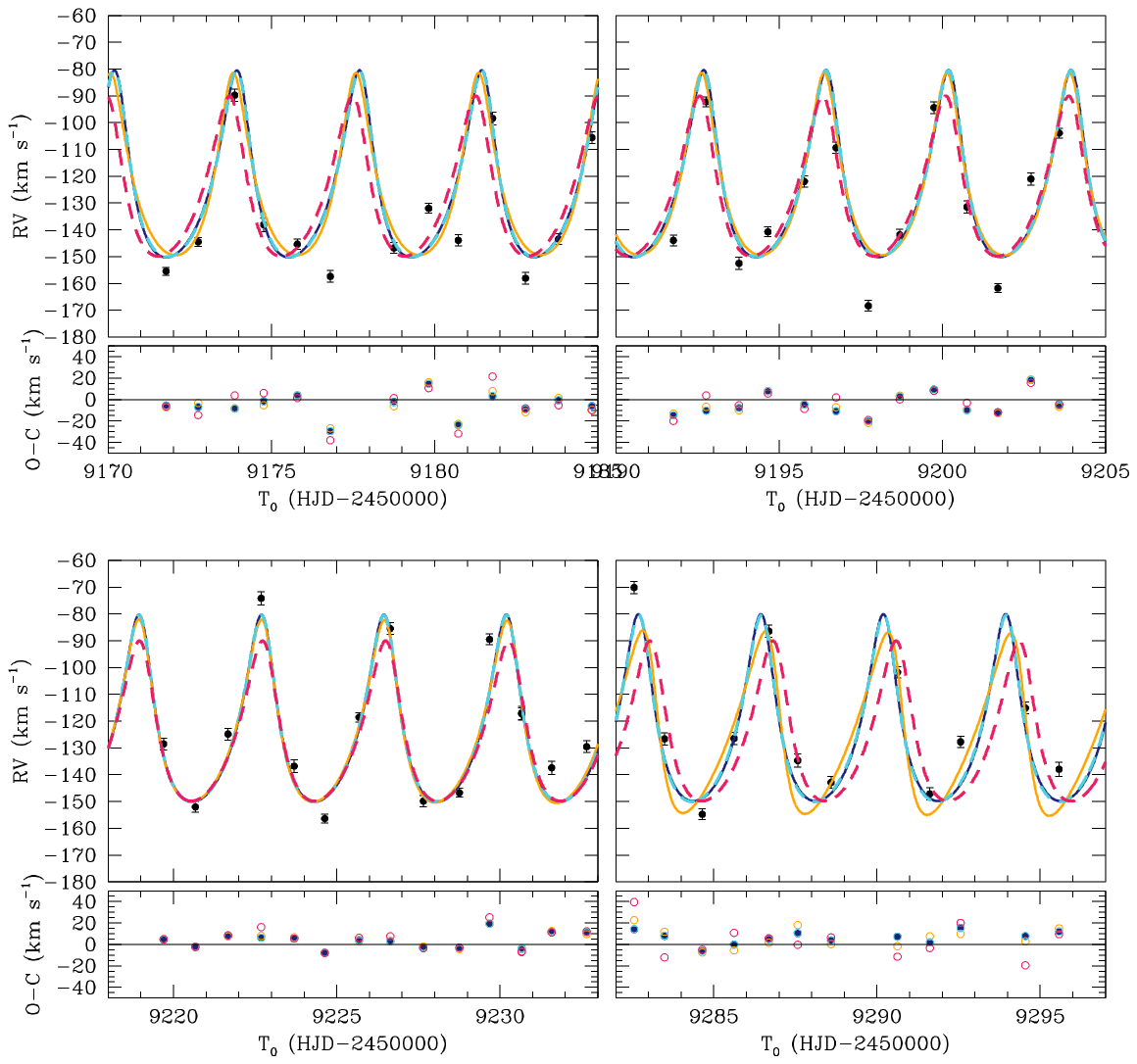}
\caption{Comparison between the measured RVs of EZ CMa (filled dots) with the orbital solutions (see Table\,\ref{table:RVs_fit}). We note that the error bars plotted on the RVs measurements are the ones quoted in Table\,\ref{Table: N V moments} multiplied by four. The dark blue plain (resp. light blue dashed) line represents the fitted RV curve with (resp. without) apsidal motion and $P_\text{orb}$ free. The orange plain (resp. pink dashed) line represents the fitted RV curve with (resp. without) apsidal motion and $P_\text{orb}$ fixed to 3.77\,days.}
\label{fig:fitRV}
\end{figure*}

We repeated the same exercise but fixing the orbital period to the values of 3.77\,days used by \citet{1997ApJ...482..470M}. The best-fit values of the adjustment are provided in the first column in Table\,\ref{table:RVs_fit} and the best-fit RV curve is  plotted in orange in Fig.\,\ref{fig:fitRV} together with the observational data and residuals of the fit. The value of the apsidal motion rate we obtain seems unreasonably high for a system where EZ CMa would be the more massive star. Indeed, this would be two orders of magnitude higher than the values found for massive O star binaries with similar periods in the LMC \citep{2020A&A...640A..33Z}.

We therefore tested the hypothesis of no apsidal motion in the system by fixing the apsidal motion rate to $0^\circ\,\text{yr}^{-1}$. The best-fit adjustment is shown in pink  in Fig.\,\ref{fig:fitRV} and the best-fit values are provided in the third column in Table\,\ref{table:RVs_fit}. The best-fit adjustment obtained in this way clearly fails to reproduce the data. This result suggests that an orbital period of 3.77\,days might not be adequate to reproduce the data.

\begin{figure*}
\centering
\hspace{-6.68cm}\includegraphics[clip=true,trim=10 20 0 200, width=0.615\linewidth]{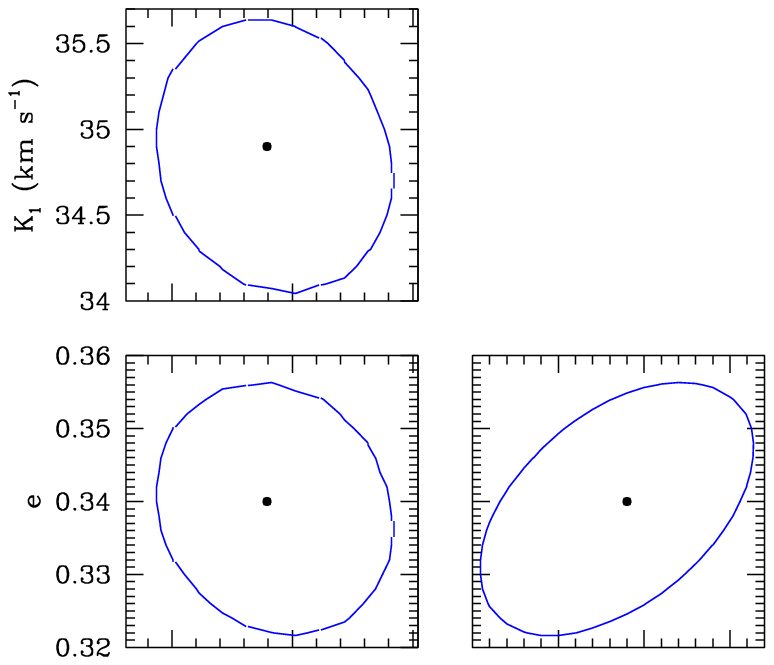}\\
\vspace{-0.5cm}
\includegraphics[clip=true,trim=10 20 0 30, width=0.615\linewidth]{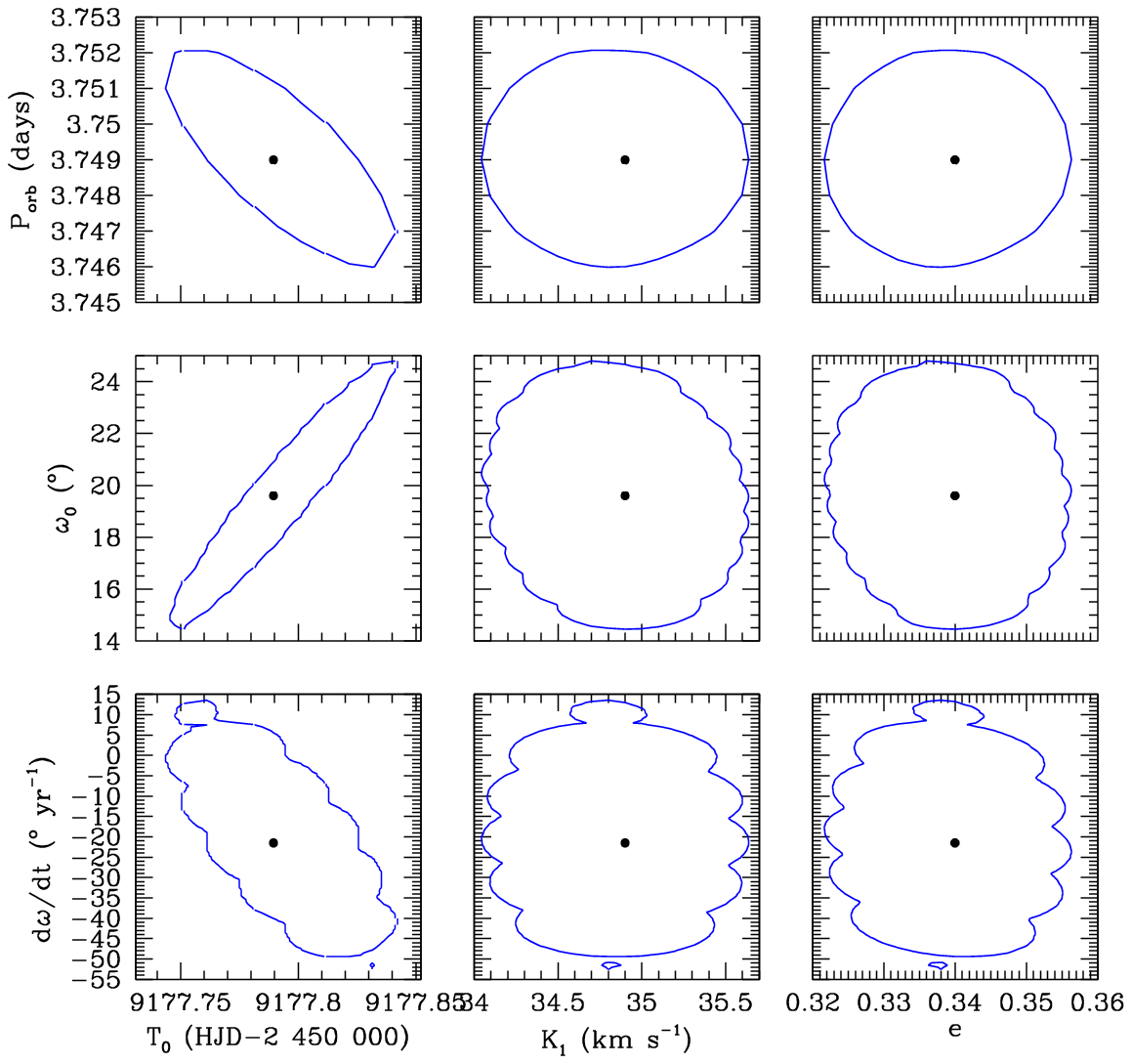}
\hspace{-0.3cm}
\includegraphics[clip=true,trim=60 20 180 200, width=0.364\linewidth]{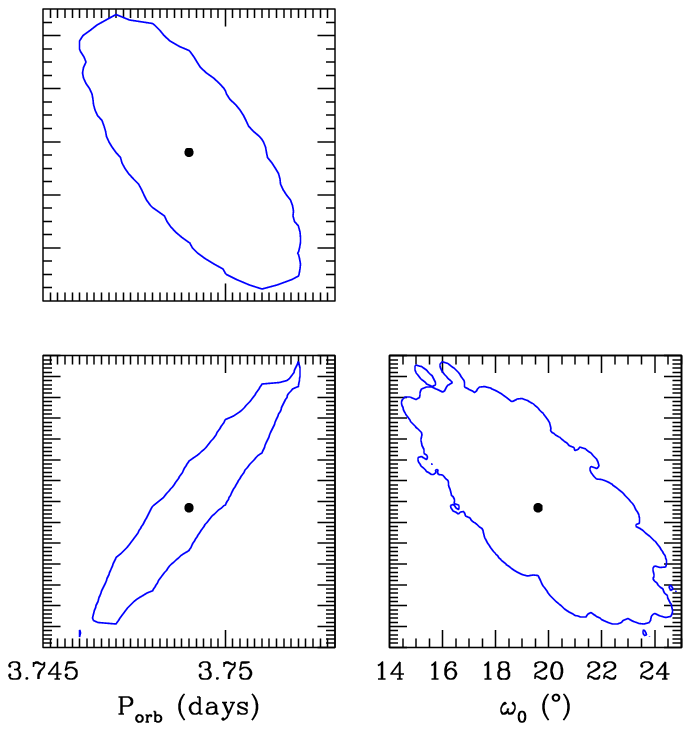}
\caption{Confidence contours for the best-fit parameters obtained from the fit of the RV data of EZ CMa allowing for apsidal motion in the system and with the orbital period a free parameter of the fit (see Column\,2 of Table\,\ref{table:RVs_fit}). The best-fit solution is shown in each panel by the black filled dot. The corresponding $1\sigma$ confidence level is shown by the blue contour.}
\label{fig:contours_RVs}
\end{figure*}

Given the strong dependence of the best-fit adjustment on the orbital period and the fact that previous orbital period determinations might not be accurate enough, we performed a last adjustment of the RV data leaving the orbital period as an additional free parameter. The projections of the 6D parameter space onto the 2D planes are illustrated in Fig.\,\ref{fig:contours_RVs}.  Our best-fit adjustment is plotted in dark blue in Fig.\,\ref{fig:fitRV} and the best-fit values are provided in the second column in Table\,\ref{table:RVs_fit}. Compared to the adjustment with $P_\text{orb}=3.77$\,days, the $\chi^2$ of the fit is not significantly better but our best-fit solution better reproduces the data at later times.

Yet, the apsidal motion rate obtained in this way is slightly negative, meaning that we have a retrograde motion. This negative apsidal motion cannot arise from the binary system alone and, if real, would be produced by a third body orbiting the inner binary \citep{2019EAS....82...99B}. However, the uncertainties on the apsidal motion are large, and the best-fit value is compatible with positive values, that is to say, with prograde apsidal motion, or no apsidal motion. These results show a degeneracy between the apsidal motion and orbital period and we are therefore unable to provide any strong determination of the apsidal motion rate in the system.

To illustrate this degeneracy, we assumed that the system undergoes no apsidal motion. We performed a last fit fixing the apsidal motion rate to $0^\circ\,\text{yr}^{-1}$. The best-fit solution is plotted in light blue in Fig.\,\ref{fig:fitRV} and overlaps perfectly with the dark blue curve; and the best-fit values are reported in the last column in Table\,\ref{table:RVs_fit}. The quality of the fit has not significantly changed compared to the previous adjustment, as judged by its $\chi^2$.

We conclude, from the analysis of the RVs of EZ CMa, that the orbital periods of 3.626\,days and 3.77\,days reported in the literature are not appropriate to explain the observed variations in RVs. If EZ CMa happened to belong to a binary system, we rather suggest an orbital period of $3.751\pm0.001$\,days to explain the variations of RVs. However, we cannot confirm nor refute the apsidal motion hypothesis based on the RVs analysed here, as our best-fit adjustment was compatible with prograde, retrograde, and no apsidal motion.

\begin{table*}
\caption{Best-fit orbital parameters of EZ CMa obtained from the fit of the RV data.}
\label{table:RVs_fit}
\centering
\begin{tabular}{l l l l l}
\hline\hline
\vspace{-3mm}\\
Parameter & \multicolumn{2}{l}{With apsidal motion} & \multicolumn{2}{l}{Without apsidal motion} \\
& $P_\text{orb}$ fixed & $P_\text{orb}$ free & $P_\text{orb}$ fixed & $P_\text{orb}$ free\\ 
\hline
\vspace*{-3mm}\\
$P_{\rm orb}$\,(d) & 3.77 (fixed) & $3.749\pm0.003$& 3.77 (fixed) & $3.751\pm0.001$ \\
\vspace*{-3mm}\\
$e$ &  $0.331^{+0.018}_{-0.019}$  & $0.340^{+0.016}_{-0.019}$ & $0.302\pm0.021$  & $0.338^{+0.018}_{-0.017}$\\ 
\vspace*{-3mm}\\
$K_{\rm P}$\,(km\,s$^{-1}$) & $34.0\pm0.8$  & $34.9^{+0.7}_{-0.9}$ & $29.9\pm0.9$ & $34.8\pm 0.8$\\
\vspace*{-3mm}\\
$T_0$ (HJD-2\,450\,000) & $9177.57^{+0.04}_{-0.03}$ & $9177.79\pm0.05$ & $9177.61\pm0.04$  & $9177.77^{+0.04}_{-0.03}$\\
\vspace*{-3mm}\\
$\omega_0$ ($^\circ$)  & $-4.7^{+3.8}_{-3.6}$ &  $19.6\pm5.2$ & $23.4\pm4.5$  & $17.1^{+3.7}_{-3.4}$\\
\vspace*{-3mm}\\
$\dot{\omega}$ ($^{\circ}$\,yr$^{-1}$) & $207.2^{+10.3}_{-10.7}$ & $-21.5^{+35.3}_{-31.3}$  & 0 (fixed) & 0 (fixed)\\
\vspace*{-3mm}\\
$\chi^2_\nu$ & 28.113  & 26.669 & 42.417  & 26.131\\
\vspace*{-3mm}\\
\hline
\end{tabular}
\end{table*}

\section{Discussion}
\label{sect:discussion}
Even if the binary hypothesis investigated in Sect.\,\ref{sect:apsidal_motion} could be a reasonable explanation to the observed RV variations, other studies and properties of EZ CMa make the binary hypothesis difficult to accept. For instance, the star was modeled in the sample of Galactic WN stars by \citet{2006A&A...457.1015H}, where the emission lines are fully reproduced in strength without any dilution from a second star, which argues against a bright companion. It was also noted that the X-ray flux is stronger than expected for a single WN star, but is also too weak for EZ CMa to be considered as a high-mass X-ray binary within the usual $L_X-L_{\rm bol}$-relation for single O stars \citep{2005MNRAS.361..679O}. This means that the putative companion should not be a black hole or a neutron star. Furthermore, the radio spectrum is observed to be thermal \citep{2000MNRAS.319.1005D}, further supporting a single-star scenario. The models of \citet{2019A&A...624L...3S} are more suggestive of a low-mass non-compact companion star, that due to its low luminosity, may not provide enough light to dilute the optical emission lines. 

\begin{figure}
\centering
\includegraphics[clip=true,trim=0 10 270 0,width=\linewidth]{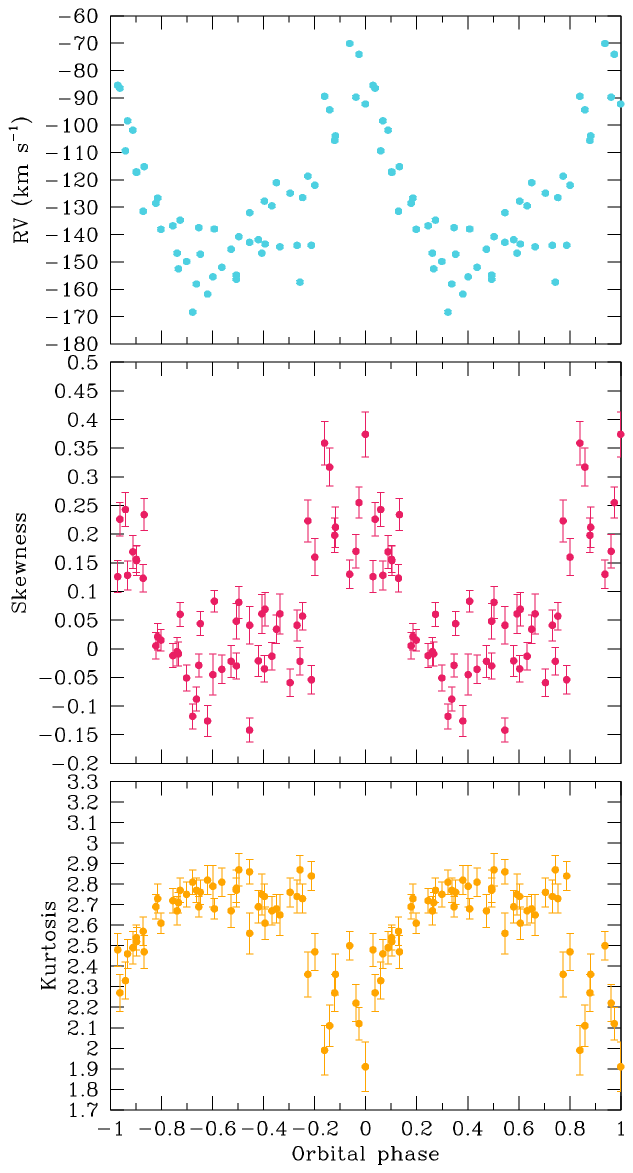}
\caption{Radial velocity (\textit{top panel}), skewness (\textit{middle panel}), and kurtosis (\textit{bottom panel}) as a function of the orbital phase computed using the orbital solution without apsidal motion and $P_\text{orb} = 3.751$\,days (see last column in Table\,\ref{table:RVs_fit}).}
\label{fig:SandK}
\end{figure}

The binary scenario suggested by \citet{2019A&A...624L...3S} and \citet{2020A&A...639A..18K} was based entirely on the 5-month long time-series of {\it BRITE-Constellation} photometry originally reported by \citet{2018pas8.conf...37M}. If we assume that the photometric signal is caused by eclipses from a companion-induced shock, referring to the scenario hypothesized by \citet{2019A&A...624L...3S} where emission from a wind-wind interaction region undergoes eclipses by the WR star and by its putative companion, then, even with the precession, this signal should always be present as the inclination of the system should remain constant over timescales of decades. \citet{1992ApJ...397..277R} compiled a collection of light-curves of the star up to that date, and the photometry was non-variable (within the errors on the data) in at least three observing runs in 1976, 1978 \citep{1980ApJ...239..607F}, and 1986 \citep{1987A&A...185..131V, 1990A&A...228..108V}. It is worth comparing that the typical scatter and errors in those data runs are smaller than the amplitude of the photometry obtained by {\it BRITE-Constellation}. Any precessing model of the binary should also be able to explain a non-variable stellar flux at these epochs, along with the various other shapes of the light curves shown in the compilation of data by \citet{1992ApJ...397..277R}.

\begin{figure}
\centering
\includegraphics[width=\columnwidth]{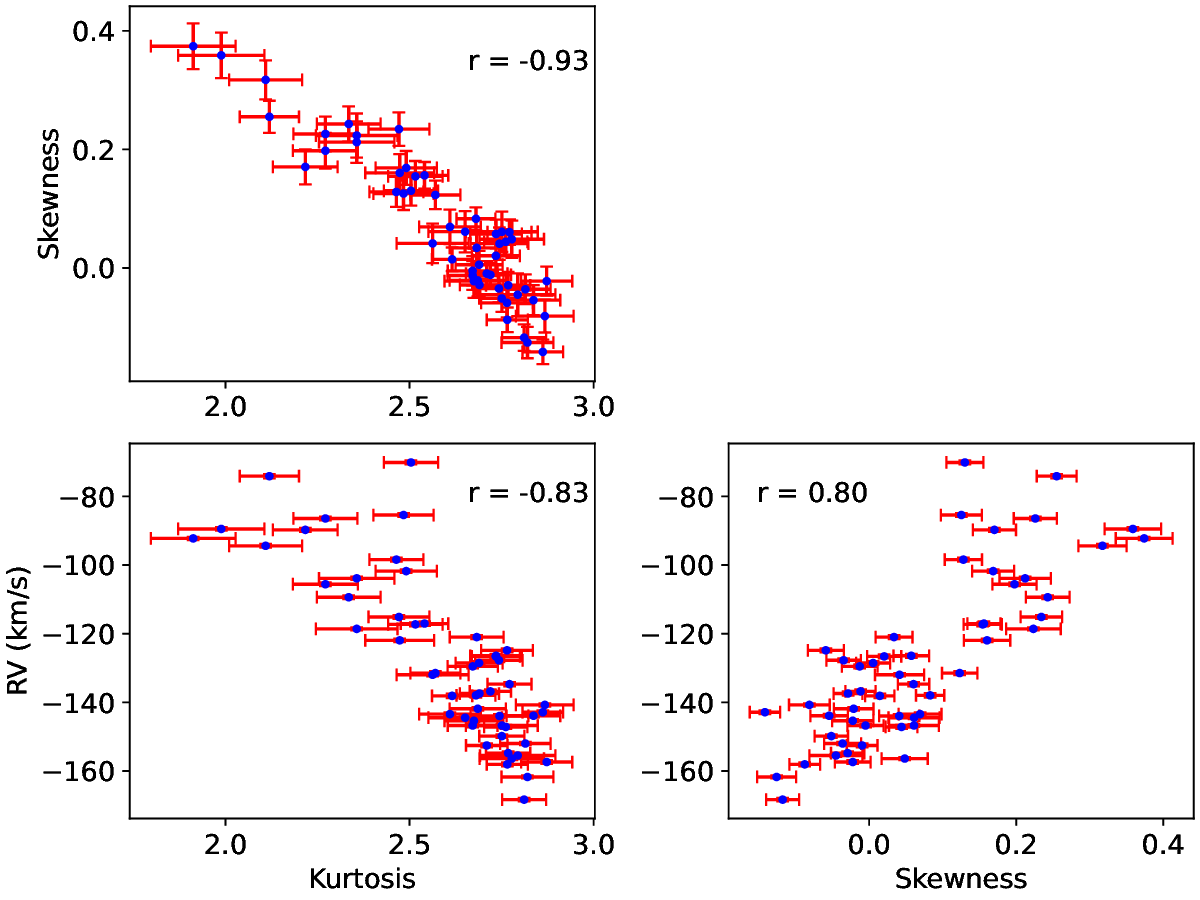}
\caption{Interdependence of the moments for the N\,{\sc v} $\lambda\,4945$ line of our observations.}
\label{fig:corner plot}
\end{figure}

Recently, EZ CMa was also considered to be a binary in the multiplicity survey of \citet{2022arXiv220412518D}. This survey relied on cross-correlation of individual epochs of spectroscopy against the average for the star from their high-resolution spectra. The normalized moments (skewness and kurtosis) of the N\,{\sc v} $\lambda\,4945$ line are shown as a function of the phase (computed using the orbital solution without apsidal motion and $P_\text{orb}=3.751$\,days, see last Column in Table\,\ref{table:RVs_fit}) along with the RVs. These plots show a strong correlation between the N\,{\sc v} $\lambda\,4945$ line's RVs as measured with a flux-weighted mean, the skewness of the line, and the kurtosis of the line. The phased variability of the line moments on the putative orbit strongly suggests that the line's intrinsic variations are the cause of the apparent RV changes more than a binary companion. We further show the correlations between the moments in Fig.\,\ref{fig:corner plot}, which all have a Spearman rank-order correlation larger than 0.8.

\begin{figure*}
\centering
\includegraphics[width=3.5in]{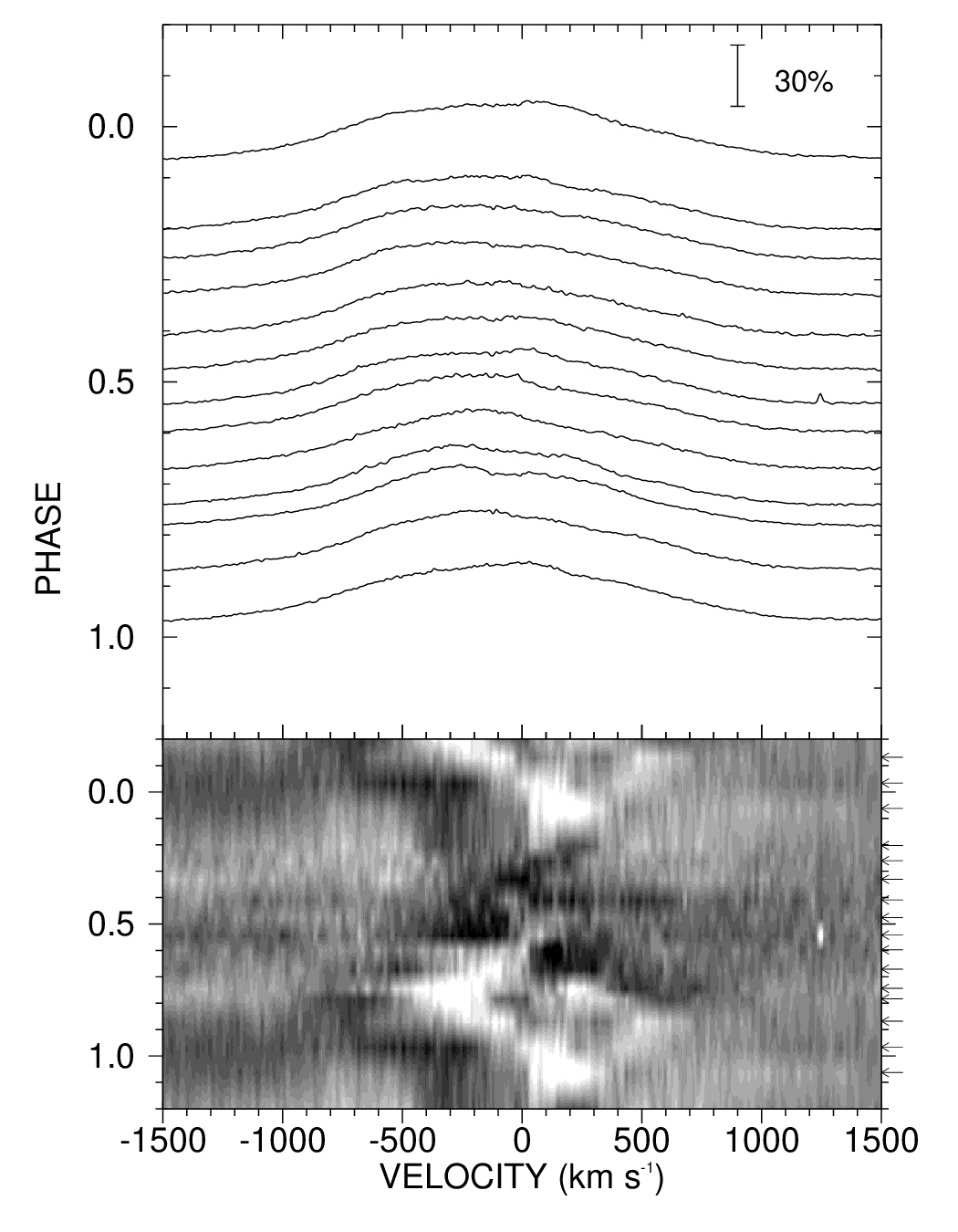}
\includegraphics[width=3.5in]{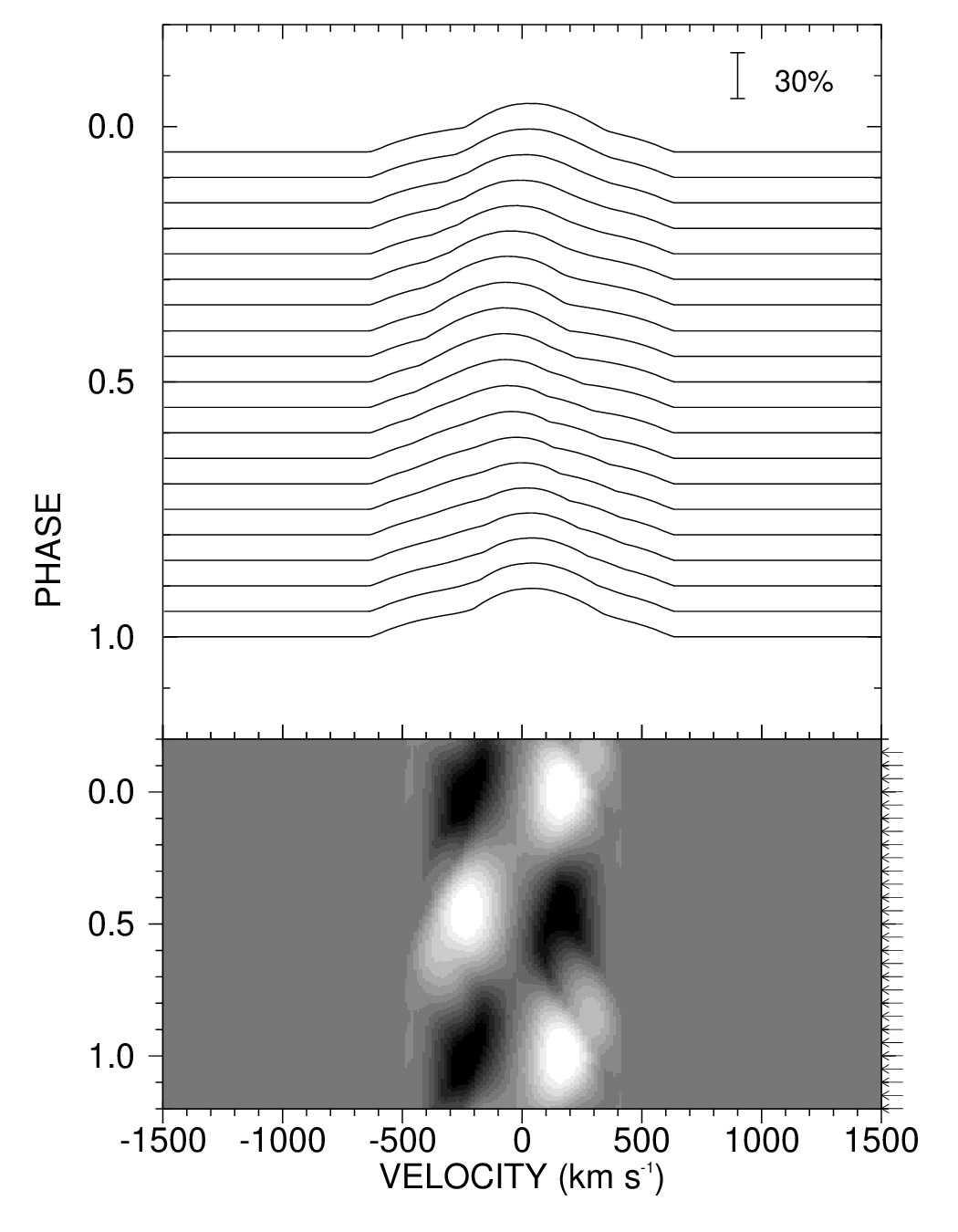}
\caption{Left panel: Section 1 (from 2020 November 17--30) of the data presented as a grey scale dynamical spectrum with line profiles. The grey scale depicts the difference from the average profile in this Section of data and are presented in the same order as the top panel. Right panel: example grey scale and model line profiles based on the models of \citet{2018MNRAS.474.1886S} shown in the same manner. }
\label{fig:cir model}
\end{figure*}

These moment variations indicate intrinsic changes within the N\,{\sc v} $\lambda\,4945$ line. If this line is variable due to a binary companion (e.g., to a moving absorption line), then a secondary star could be the cause of the observed changes. This would require a very hot and luminous underlying star with an underlying N\,{\sc v} $\lambda\,4945$ line. We note that even the very bright and hot O4\,I star $\zeta$ Puppis does not show any strong enough N\,{\sc v} $\lambda\,4945$ absorption line \citep[see the average high-resolution CHIRON spectrum in][]{2018MNRAS.473.5532R} to provide such a variation, so any object causing the N\,{\sc v}  $\lambda\,4945$ variability would have to have an extremely hot temperature to cause these changes in the spectrum of the WN4 star EZ CMa. Such a star should easily be observable in a combined spectrum and cause diluted emission lines. 

Many observational results presented both here and in the literature call into question the binary hypothesis, including the correlation of line moments with radial velocity, occasional non-variable photometry in the past, and the result here that only the N\,{\sc v} $\lambda\,4945$ line was able to be used for a binary orbital solution. One of the best samples of precessing, massive binaries was presented in \citet{2020A&A...640A..33Z}, who analyzed 162 binaries in the LMC with OGLE photometry. While the putative binary period of EZ CMa is similar to the periods shown in this sample, the apsidal period proposed by \citet{2019A&A...624L...3S} and \citet{2020A&A...639A..18K} is about two orders of magnitude shorter than most of the massive binaries in the sample of \citet{2020A&A...640A..33Z}, although one outlier exists with similar parameters. 

The fast apsidal motion rates derived for an orbital period of 3.77\,days, if real, raises several issues.
Indeed, in such a case, the apsidal motion cannot be considered as a small secular variation anymore, and the orbital fit performed in this case is likely not relevant.
With such a large value for the apsidal motion rate, the remaining questions relate to the tidal force dissipation in the binary. For late-type binaries in young clusters, tidal forces cause circularization quickly, as shown in e.g., \citet{1995ApJ...444..338T}, \citet{1997ApJ...474..760R}, and \citet{1997ApJ...481..363T}. While the exact timescales and mechanisms are debated in these papers, the timescales are still seen to be short, and would likely be only $10^{3-4}$ yr for a system with two massive stars such as the expected types for WR6, which does not incorporate potential changes due to the stellar structure of the WR star being different from that of a (more centrally compressed) main-sequence star.

The difficulties in the binary scenario suggest that we should examine these data with an alternative hypothesis such as CIRs, which were previously suggested for EZ CMa by \citet{1997ApJ...482..470M}. CIRs have been long-observed in O stars \citep{1984ApJ...283..303M}, and are  observed as propagating absorption features in ultraviolet resonance lines \citep{1999A&A...344..231K} and are ubiquitous amongst the O stars \citep[e.g.,][]{1989ApJS...69..527H}. 

CIRs have been harder to detect and characterize for WR stars. \citet{2009ApJ...698.1951S} and \citet{2011ApJ...736..140C} performed optical spectroscopic surveys of WR stars to search for line profile variability and test the strength of the variations for quantifying CIRs. Prior to this, \citet{1997ApJ...482..470M, 1999ApJ...518..428M} found EZ CMa and WR 134 to have line profile variations similar to that expected for CIRs. The surveys found several new candidates for CIR-hosting WN stars, including WR 1 \citep{2010ApJ...716..929C} and WR 110 \citep{2011ApJ...735...34C}, which also showed photometric variations with the \textit{Microvariability and Oscillations of STars} (\textit{MOST}) satellite. 

EZ CMa has some similarities in terms of spectroscopic features with another well-studied WR star with CIRs, namely WR 134 \citep{1999ApJ...518..428M}. For WR 134, \citet{2016MNRAS.460.3407A} collected four months of moderate-resolution spectroscopy for their analysis. They examined the variability patterns and derived a 40-d lifetime for the features. WR 134 has similar X-ray and optical properties to EZ CMa, so it is reasonable to assume they have similar causes of their variations. With EZ CMa, the coherence time of the cyclic variations is about two weeks. While we can't directly confirm this with our data, the different variability patterns in each epoch do seem to support this.

CIRs in a WR wind could provide shock velocities up to $10^3$\,km\,s$^{-1}$ assuming energy conservation between thermal and kinetic energy. These shock velocities are in the radial direction given the terminal wind speeds of these stars, allowing the stars to also exhibit a fairly hard X-ray flux, such as observed in EZ CMa \citep{2012ApJ...747L..25O,2015ApJ...815...29H}. This is a higher shock velocity than the shocks from the clumps in the winds \citep[$\sim 10^2\,\text{km}\,\text{s}^{-1}$][]{1999ApJ...514..909L}. Thus, CIRs can provide the needed properties to explain the observed properties of EZ CMa. In fact, \citet{2012ApJ...747L..25O} show that the X-ray production for EZ CMa is difficult to reconcile with a binary hypothesis and that the X-ray production is not necessarily consistent with a normal wind, and offers a slowly accelerated clump as a potential source of the X-rays. These clumps could be part of the base of the CIRs in the wind. Similar to EZ CMa, 40\% of supposedly single WN stars show similar optical spectroscopic variability \citep{2011ApJ...736..140C}. Two O stars have been shown to have CIRs in both ultraviolet and optical spectroscopy along with high-precision optical photometry, namely $\xi$ Per \citep{2014MNRAS.441..910R} and $\zeta$ Pup \citep{2018MNRAS.473.5532R} which also shows X-ray variations at the 10 -- 20\% level \citep{2021ApJ...906...89N}. It is worth noting that X-ray production in CIRs is not well modeled and could cause excess emission or absorption which may depend on the source and geometry, so it is unclear how the X-rays could relate to the CIRs without further modeling, which is beyond the scope of this paper.

Modeling of the CIRs is beginning to be explored with multiple observational techniques. Radio emission from CIRs may show modulation \citep{2020MNRAS.497.1127I} and it is likely that similar variability patterns could be observed with optical photometry although a radio photosphere may lie outside the area where the CIRs form. Simplified numerical simulations based on the results of \citet{1996ApJ...462..469C} and using Monte Carlo radiative transfer predicts the shape of light and polarization curves of winds that include CIRs \citep{2019MNRAS.489.2873C}. 

For the current observations the CIR models rely on a hot spot or a pair of hot spots that are at an angle of 90$^\circ$ from each other on the stellar surface acting as the source(s) of the large-scale wind inhomogeneities that cause the CIR. The resulting patterns of variability modulate with the underlying star with a period independent of the distance from the star. 
The polarimetric variability of EZ CMa was modeled in the context of CIRs by \citet{2018MNRAS.474.1886S} using the analytical results of \citet{2015A&A...575A.129I}, and the models of \citet{2018MNRAS.474.1886S} are the basis for our calculations. These models are generally set up to have hot spots as drivers for the CIRs, and were able to reproduce many epochs of polarimetric variability by only changing the location of the footprint of the CIRs on the stellar surface of the star as a function of epoch. In addition to the success of the polarimetric variability, the study also produced generic line profiles that could be used to examine spectroscopic signatures of CIRs. 

In Fig.\,\ref{fig:cir model}, we present a graphical depiction of the line profiles from Section 1 of our data from 2020 November 17--30, both showing a montage of the line profiles as well as a grey scale of the difference from each spectrum and the average profile from that section. A clear moving excess emission can be seen to propagate around the rest velocity of the line with the 3.77 day period. A similar result was obtained by \citet[][see their Fig.\,4]{1998ApJ...498..413M} for this transition, although they also reported on several additional lines. As a qualitative comparison, we also show a model of line profile variability for two CIRs with the code developed by \citet{2018MNRAS.474.1886S}. The model looks out of phase compared to our observations as it was not developed to fit these data, but shows a qualitative similarity to the observed variations. This model can illustrate several plausible similarities with the data although a full model of the system and its CIRs is beyond the scope of this paper and should be done with a denser data set than we obtained.

The model line profile used for Fig.\,\ref{fig:cir model} follows the outline of section 5 of \cite{2018MNRAS.474.1886S} for the calculation of recombination emission lines from a wind with a CIR.  The approach uses the Sobolev approximation for the line formation \citep[e.g.,][]{1999isw..book.....L}, where the emissivity for recombination scales as the square of density.  The emissivity also involves a function $(v/v_\infty)^\alpha$ for $v$ the wind velocity and $\alpha$ a user-supplied exponent to shift the line emission to larger velocities, helping the profile to match the observed line width better.  Then two models are calculated for the line profile variations, each with a single CIR but at different azimuthal locations on the equator.  Those results are averaged to obtain the results shown in Fig.\,\ref{fig:cir model}. Since the observed N\,{\sc v} $\lambda\,4945$ variations in the profile do not reach $v_\infty$ due to the high ionization of this line, we scaled the velocities by a factor of 3 in this depiction. The azimuthal separation of the two CIRs is about $90^\circ$ \citep[following][]{2018MNRAS.474.1886S}, and the half-opening angles are $30^\circ$ each, so in the model the two CIRs nowhere physically intersect.  Thus the averaging of the results from the separate models for the individual CIRs is a reasonable approximation for the pattern of line profile modulations with rotational phase.

The model shown in the grey scales (Fig.\,\ref{fig:cir model}) produces variations in the moments that are of similar amplitude or shape as the measured moments from our data. However, the line profiles of the CIR model are narrower than that of the WR star due to the method by which the simplified profile is calculated. We measured the normalized moments of the model profiles in the same manner as our data. We found that the observed variations are similar in both amplitude and shape as that of the model. 

We do caution that these line profiles were calculated with some assumptions that could make this a difficult comparison to ultraviolet resonance lines formed by CIRs, where the observational and modeled effect is a ``banana'' shape in the dynamical spectra \citep[e.g., ][]{1996ApJ...462..469C}. In particular, optical recombination lines are sensitive to the squared density in the wind rather than the velocity gradients that drive the observed ``banana'' shape for the P\,Cygni absorption profiles of ultraviolet resonance lines. These line profiles do not take into account the non-monotonic velocity gradient effects, but are presumably a realistic first step towards understanding the CIR effects.

Of note, we also consider if this pattern can be caused by binarity, but we caution that for binarity, the time-scale must be on the orbital period strictly, whereas rotation will have a cadence that depends on rotation period and number of spots. The changes in the variability pattern are thus hard to reconcile with binarity. Further, the WR wind is spherical up to the stagnation point, which is likely close to the companion for a system like EZ CMa. If we are looking at a line that forms at the inner wind of the WR star such as N V, then the companion might reside at a position similar to where the WR wind reaches its terminal speed. The spiral trail effects could then influence line wings and could begin to constrain viewing inclination combined with the CIR opening angle. This has been explored in a paper on forbidden lines for WR binaries \citep{2009MNRAS.395..962I}, which is a different scale of the problem, but has all relevant issues covered. Finally, our CIRs are mostly conical with little spirality in the region of line formation, especially for N V. Thus, spiral patterns in the dynamical spectra do not imply a spiral pattern in the spatial structure, as it is a fully 3D problem with spherical trig considerations to map from spatial to observed radial velocity.

\section{Conclusion}
\label{sect:conclusion}
Our spectroscopic campaign on EZ CMa aimed at testing the hypothesis that the photometric variations observed with the {\it BRITE-Constellation} nanosatellites are caused by a fast precession in a close binary system. We focused our analysis on the sole N\,{\sc v} $\lambda\,4945$ line and fitted the radial velocities of EZ CMa explicitly accounting for apsidal motion in the system. 

We concluded that the orbital period of 3.626\,days proposed by \citet{2019A&A...624L...3S} is not appropriate to describe the observed RV variations. We did find an orbital solution with the period of 3.77\,days used by \citet{1997ApJ...482..470M}, but at the cost of an extremely fast apsidal motion of $207.2^\circ\,\text{yr}^{-1}$, unreasonably high for a system where EZ CMa would be the more massive star. 
We rather suggest an orbital period of $3.751\pm0.001$\,days to explain the observed variations in RVs, if EZ CMa would appear to belong to a binary system. Our best-fit orbital solution is compatible with no apsidal motion, which means that we cannot confirm nor deny the hypothesis of precession of the orbit based on the present analysis only.

In addition, the radial velocity changes alone are not enough to consider the system to be a binary. The system had periods of non-variable photometry, no emission line dilution, and any companion that might be causing the changes in the N\,{\sc v} $\lambda\,4945$ line profile should lead to a luminosity of the companion that would be detectable by our spectroscopy or by past campaigns. The proposed companion from the analyses of \citet{2019A&A...624L...3S} and \citet{2020A&A...639A..18K} is a late B star with 3--5\,$M_\odot$, which should have a luminosity of $\sim300\,L_\odot$, well below the nominal detection limits. However, if the star did have a N\,{\sc v} $\lambda4945$ absorption line that could explain the skewness and kurtosis variations in this line, it should be closer to that of a mid-O dwarf and would be easily detectable for the existing spectroscopy of EZ CMa given the similar luminosities of these two types of stars. 

Instead of the binary system hypothesis, we suggest that EZ CMa is a single WR star with strong co-rotating interaction regions. Such a system is also seen in the well-studied WR star WR 134 \citep[e.g.,][]{2016MNRAS.460.3407A} and represents a simpler interpretation of all available observations. It would be prudent to have a dedicated, high-precision, high-cadence time series of spectroscopy along with photometry and polarimetry in the future to model the CIRs to definitely confirm or refute the binary hypothesis.

\section*{Acknowledgements}
This research is based on observations obtained through NOIR Lab with program NOIRLab-20A-0054. This research has used data from the CTIO/SMARTS 1.5m telescope, which is operated as part of the SMARTS Consortium by RECONS (www.recons.org) members Todd Henry, Hodari James, Wei-Chun Jao, and Leonardo Paredes. At the telescope, observations were carried out by Roberto Aviles and Rodrigo Hinojosa. KDGB acknowledges support from the Arizona Space Grant Program through the Embry-Riddle Aeronautical University Undergraduate Research Institute. This research was partially supported through the Embry-Riddle Aeronautical University’s Faculty Innovative Research in Science and Technology (FIRST) Program. AFJM and NSL are grateful to NSERC (Canada) for financial assistance. The authors gratefully thank the referee for their suggestions and comments towards the improvement of the manuscript.

\section*{Data Availability}
The raw spectra for this project are archived at the NOIR Lab archive. Reasonable requests for the reduced spectra will be accommodated by the corresponding author. Measurements from these data are given in Table \ref{Table: N V moments}.

\bibliographystyle{mnras}
\bibliography{WorkCite.bib} 

\clearpage
\appendix
\section{Moments for the N\,{\sc v} $\lambda\,4945$ Line}
\label{sect:moments}
This appendix provides the moments for the N\,{\sc v} $\lambda\,4945$ line (see Table\,\ref{Table: N V moments}). 
\begin{table}
\caption{Measurements of the moments for the N\,{\sc v} $\lambda\,4945$ line.}
\label{Table: N V moments}
\centering
\begin{tabular}{cccc}
Observation date (HJD) & RV (km/s) & Skewness & Kurtosis \\
\hline 
    2459171.7721 & $-155.48\pm0.37$ & $-0.045\pm0.036$  & $2.79\pm0.10$ \\
    2459172.7597 & $-144.47\pm0.41$ & $0.061\pm0.035$   & $2.65\pm0.10$ \\
    2459173.8727 & $-89.76\pm0.570$  & $0.170\pm0.029$   & $2.22\pm0.09$ \\
    2459174.7644 & $-138.10\pm0.64$ & $0.015\pm0.019$   & $2.61\pm0.05$ \\
    2459175.7938 & $-145.36\pm0.49$ & $-0.022\pm0.028$  & $2.67\pm0.08$ \\
    2459176.8036 & $-157.38\pm0.55$ & $-0.022\pm0.024$  & $2.87\pm0.07$ \\
    2459178.7516 & $-146.75\pm0.51$ & $-0.005\pm0.025$  & $2.67\pm0.07$ \\
    2459179.8136 & $-131.98\pm0.46$ & $0.041\pm0.033$   & $2.56\pm0.10$ \\
    2459180.7236 & $-143.91\pm0.53$ & $-0.054\pm0.025$  & $2.84\pm0.07$ \\
    2459181.7758 & $-98.43\pm0.590$  & $0.128\pm0.025$   & $2.46\pm0.07$ \\
    2459182.7861 & $-158.06\pm0.56$ & $-0.088\pm0.021$  & $2.77\pm0.06$ \\
    2459183.7926 & $-143.43\pm0.49$ & $0.069\pm0.029$   & $2.61\pm0.08$ \\
    2459184.8136 & $-105.60\pm0.53$ & $0.198\pm0.030$   & $2.27\pm0.09$ \\
    2459191.7682 & $-143.97\pm0.52$ & $0.041\pm0.027$   & $2.74\pm0.08$ \\
    2459192.7714 & $-92.25\pm0.470$  & $0.374\pm0.039$   & $1.91\pm0.12$ \\
    2459193.7741 & $-152.55\pm0.57$ & $-0.009\pm0.021$  & $2.71\pm0.05$ \\
    2459194.6586 & $-140.74\pm0.46$ & $0.081\pm0.028$   & $2.87\pm0.08$ \\
    2459195.7780 & $-121.94\pm0.51$ & $0.160\pm0.032$   & $2.47\pm0.09$ \\
    2459196.7470 & $-109.39\pm0.53$ & $0.243\pm0.030$   & $2.33\pm0.09$ \\
    2459197.7327 & $-168.33\pm0.51$ & $-0.118\pm0.022$  & $2.81\pm0.06$ \\
    2459198.6975 & $-141.86\pm0.50$ & $-0.021\pm0.027$  & $2.69\pm0.08$ \\
    2459199.7483 & $-94.41\pm0.550$  & $0.317\pm0.033$   & $2.11\pm0.10$ \\
    2459200.7617 & $-131.47\pm0.58$ & $0.123\pm0.024$   & $2.57\pm0.07$ \\
    2459201.7040 & $-161.73\pm0.42$ & $-0.126\pm0.027$  & $2.82\pm0.07$ \\
    2459202.7124 & $-120.99\pm0.57$ & $0.034\pm0.025$   & $2.68\pm0.07$ \\
    2459203.5833 & $-103.88\pm0.45$ & $0.212\pm0.035$   & $2.36\pm0.10$ \\
    2459219.7010 & $-128.54\pm0.54$ & $0.005\pm0.023$   & $2.69\pm0.06$ \\
    2459220.6677 & $-151.99\pm0.48$ & $-0.036\pm0.025$  & $2.81\pm0.07$ \\
    2459221.6715 & $-124.84\pm0.55$ & $-0.059\pm0.024$  & $2.76\pm0.07$ \\
    2459222.6851 & $-74.13\pm0.620$  & $0.255\pm0.027$   & $2.12\pm0.08$ \\
    2459223.6970 & $-136.80\pm0.59$ & $-0.012\pm0.021$  & $2.72\pm0.06$ \\
    2459224.6308 & $-156.37\pm0.41$ & $0.048\pm0.031$   & $2.78\pm0.09$ \\
    2459225.6850 & $-118.59\pm0.45$ & $0.223\pm0.037$   & $2.36\pm0.11$ \\
    2459226.6388 & $-85.44\pm0.550$  & $0.126\pm0.028$   & $2.48\pm0.08$ \\
    2459227.6515 & $-149.83\pm0.53$ & $-0.051\pm0.023$  & $2.75\pm0.06$ \\
    2459228.7566 & $-146.75\pm0.41$ & $0.061\pm0.034$   & $2.75\pm0.10$ \\
    2459229.6796 & $-89.49\pm0.490$  & $0.359\pm0.038$   & $1.99\pm0.12$ \\
    2459230.6632 & $-117.05\pm0.63$ & $0.156\pm0.023$   & $2.54\pm0.06$ \\
    2459230.6680 & $-117.29\pm0.55$ & $0.154\pm0.026$   & $2.52\pm0.07$ \\
    2459231.5847 & $-137.42\pm0.62$ & $-0.029\pm0.020$  & $2.69\pm0.05$ \\
    2459232.6535 & $-129.52\pm0.55$ & $-0.013\pm0.024$  & $2.67\pm0.07$ \\
    2459282.5630 & $-70.13\pm0.560$  & $0.130\pm0.025$   & $2.50\pm0.07$ \\
    2459283.4950 & $-126.62\pm0.56$ & $0.021\pm0.023$   & $2.73\pm0.07$ \\
    2459284.6465 & $-154.77\pm0.50$ & $-0.030\pm0.023$  & $2.77\pm0.06$ \\
    2459285.6174 & $-126.44\pm0.58$ & $0.057\pm0.024$   & $2.73\pm0.07$ \\
    2459286.6896 & $-86.44\pm0.560$  & $0.226\pm0.029$   & $2.27\pm0.09$ \\
    2459287.5751 & $-134.70\pm0.60$ & $0.060\pm0.021$   & $2.77\pm0.06$ \\
    2459288.5935 & $-142.87\pm0.54$ & $-0.142\pm0.021$  & $2.86\pm0.06$ \\
    2459290.6284 & $-101.78\pm0.52$ & $0.169\pm0.029$   & $2.49\pm0.08$ \\
    2459291.6186 & $-147.16\pm0.57$ & $0.044\pm0.021$   & $2.76\pm0.06$ \\
    2459292.5639 & $-127.74\pm0.52$ & $-0.035\pm0.023$  & $2.74\pm0.06$ \\
    2459294.5478 & $-115.13\pm0.55$ & $0.234\pm0.028$   & $2.47\pm0.08$ \\
    2459295.5774 & $-137.96\pm0.66$ & $0.083\pm0.019$   & $2.68\pm0.05$ \\ 
    \hline
    \end{tabular}
\end{table}

\bsp	
\label{lastpage}

\end{document}